\documentstyle[aps,preprint,eqsecnum,floats,epsf]{revtex}

\tighten

\draft

\begin{document}

\preprint{} \title{An Algebraic $q$-Deformed Form for
Shape-Invariant Systems}
 
\author{A.~N.~F.       Aleixo\thanks{Electronic    address:       {\tt
armando@if.ufrj.br}}} \address{Instituto de F\'{\i}sica,  Universidade
Federal do   Rio     de  Janeiro,   RJ   -    Brazil}
\author{A.~B.  Balantekin\thanks{             Electronic
address:     {\tt  baha@nucth.physics.wisc.edu}}}
\address{Department   of    Physics,  University of Wisconsin,
Madison, Wisconsin 53706 USA}   
\author{M.~A.  C\^andido
Ribeiro\thanks{Deceased}} \address{Departamento de F\'{\i}sica  -
Instituto de Bioci\^encias, Letras  e Ci\^encias Exatas\\ UNESP, S\~ao
Jos\'e do Rio Preto, SP - Brazil}  \date{\today}   \maketitle 
\begin{abstract} 
  A quantum deformed theory applicable to all shape-invariant
  bound-state  systems is introduced by defining $q$-deformed ladder
  operators. We show these new ladder operators satisfy new
  $q$-deformed commutation relations. 
  In this context we construct an alternative
  $q$-deformed model that preserve   the shape-invariance property
  presented by primary system. q-deformed generalizations of Morse,
  Scarf, and Coulomb potentials are given as examples. 
\end{abstract} 
 
 
\date{\today} 
 
\newpage 
 
\section{Introduction}

Supersymmetric quantum mechanics \cite{Witten:nf,Cooper:1994eh} 
is generally studied in the context 
of one-dimensional systems. The partner Hamiltonians 
\begin{equation} 
\label{eqh12} 
\hat H_{_1} = \hbar\Omega\,\hat A^\dagger\hat A  
\qquad\qquad {\rm and}\qquad\qquad 
\hat H_{_2} = \hbar\Omega\,\hat A\hat A^\dagger 
\end{equation} 
are most readily written in terms of one-dimensional operators 
\begin{equation} 
\label{eqaa+} 
\hat A \equiv {1\over \sqrt{\hbar\Omega}}\left(W(x) +
\frac{i}{\sqrt{2m}}\hat p \right)  \qquad\qquad {\rm and}\qquad\qquad
\hat A^\dagger \equiv {1\over\sqrt{\hbar\Omega}}\left(W(x)  -
\frac{i}{\sqrt{2m}}\hat p\right) 
\end{equation} 
where $\hbar \Omega$ is a constant energy scale factor, introduced to
permit working with   dimensionless quantities, and $W(x)$ is the
superpotential.  A number of such pairs of Hamiltonians share an
integrability condition called shape invariance
\cite{Gendenshtein:vs}. The shape-invariance condition has an
underlying algebraic structure \cite{Balantekin:1997mg}. Note that the
definition in Eq. (\ref{eqh12}) differs from the one given in Ref. 
\cite{Balantekin:1997mg} as in the present paper all operators are
defined to be dimensionless. 

The development of quantum groups and quantum  algebras motivated
great interest in $q$-deformed algebraic structures
\cite{Drinfeld:in,Jimbo:1985vd,Jimbo:mc}, and in particular in the
$q$-harmonic oscillators. Until now quantum groups have found
applications in solid state physics \cite{Batchelor:1989uk}, nuclear
physics \cite{Barbier:1993dc,Bonatsos:1999xj}, quantum optics
\cite{Chaichian:1990ic} and in conformal field theories
\cite{Alvarez-Gaume:1988vr}.  Quantum algebras are
deformed versions of the usual Lie algebras obtained by introducing a
deformation parameter $q$. In the limit of $q$ going to the unity
these quantum algebras reduce to the usual Lie algebra.  The quantum
algebras provide us with a class of symmetries which is richer than
the usual class of Lie symmetries, the latter is contained in the
former as a special case  (when $q\rightarrow 1$). Quantum algebras
may turn out to be appropriate tools for describing symmetries of
physical systems which cannot be described  by the ordinary Lie
algebras.  
 
In this paper we present a quantum deformed generalization for all
shape-invariant   systems and show that is possible to introduce
generalized ladder operators for these systems. The paper is 
organized in the
following way, in Section II we present the algebraic formulation of 
shape invariance; in Section III we introduce the fundamental
principles of our quantum deformed generalization; in Section IV we
introduce the shape-invariant formulation for quantum deformed systems
and work out  some examples. Finally, conclusions and brief remarks
close the paper in Section V.

 
\section{Algebraic Formulation to Shape Invariance} 
 
The Hamiltonian $\hat H_{_1}$ of Eq.~(\ref{eqh12}) is called 
shape-invariant if the condition 
\begin{equation} 
\hat A(a_{_1}) \hat A^\dagger(a_{_1}) =\hat A^\dagger (a_{_2})  
\hat A(a_{_2}) + R(a_{_1}) \,, 
\label{eqsi} 
\end{equation} 
is satisfied \cite{Gendenshtein:vs}. In this equation $a_{_1}$ and
$a_{_2}$ represent 
parameters of the Hamiltonian. The parameter $a_{_2}$ is a function of 
$a_{_1}$ and the remainder $R(a_{_1})$ is independent of the dynamical 
variables such as position and momentum. As it is written the 
condition of Eq.~(\ref{eqsi}) does not require the Hamiltonian to be 
one-dimensional, and one does not need to choose the ansatz of 
Eq.~(\ref{eqaa+}). In the cases studied so far the parameters $a_{_1}$ 
and $a_{_2}$ are either related by a translation
\cite{Balantekin:1997mg,ref22} or a  
scaling \cite{Balantekin:1998wj,Aleixo:2002sa,Khare:gg}. Introducing
the similarity transformation    
that replaces $a_{_1}$ with $a_{_2}$ in a given operator  
\begin{equation} 
\hat T(a_{_1})\, \hat O(a_{_1})\, \hat T^\dagger(a_{_1}) = \hat
O(a_{_2})  
\label{eqsio} 
\end{equation} 
and the operators 
\begin{equation} 
\hat B_+ =  \hat A^\dagger(a_{_1})\hat T(a_{_1}) 
\qquad\qquad {\rm and}\qquad\qquad 
\hat B_- =\hat B_+^\dagger =  \hat T^\dagger(a_{_1})\hat A(a_{_1})\,, 
\label{eqbpm} 
\end{equation} 
the Hamiltonians of Eq.~(\ref{eqh12}) take the forms 
\begin{equation} 
\hat H_{_1} =\hbar\Omega\,\hat B_+\hat B_- 
\qquad\qquad {\rm and}\qquad\qquad 
\hat H_{_2} = \hbar\Omega\,\hat T\hat B_-\hat B_+\hat T^\dagger\,. 
\label{eqhb12} 
\end{equation} 
With Eq.~(\ref{eqsi}) one can also easily prove the commutation 
relation \cite{Balantekin:1997mg}
\begin{equation} 
[\hat B_-,\hat B_+] =  \hat T^\dagger(a_{_1})R(a_{_1})\hat T(a_{_1})
\equiv  R(a_{_0})\,,  
\label{eqcb1} 
\end{equation} 
where we used the identity 
\begin{equation} 
R(a_n) = {\hat T}(a_{_1})\,R(a_{n-1})\,{\hat T}^\dagger (a_{_1})\,, 
\label{eqran} 
\end{equation} 
valid for any $n$. Eq. (\ref{eqcb1}) suggests that $\hat{B}_-$ and
$\hat{B}_+$   
are the appropriate creation and annihilation operators for the
spectra of the shape-invariant potentials provided that their
non-commutativity with $R(a_{_1})$ is taken into account. Indeed using
the relations   
\begin{equation} 
\label{eqrbpm} 
R(a_n) \hat{B}_+ = \hat{B}_+ R(a_{n-1}) \qquad\qquad {\rm
  and}\qquad\qquad  
R(a_n) \hat{B}_- = \hat{B}_+ R(a_{n+1}),  
\end{equation} 
which readily follow from Eqs. (\ref{eqbpm}) and (\ref{eqran}), one
can write down the additional commutation relations  
\begin{eqnarray} 
\label{eqhbn} 
\cases{[\hat{B}_+, R(a_{_0}) ] = \left\{R(a_{_1})-R(a_{_0})\right\} 
\hat{B}_+ \cr 
[\hat{B}_+, \left\{R(a_{_1})-R(a_{_0})\right\}] = 
\left\{R(a_{_2})-R(a_{_0})\right\}\hat{B}_+^2\,,\cr} 
\end{eqnarray} 
and the Hermitian conjugate relations. In general there is an 
infinite number of such commutation relations, hence the appropriate 
Lie algebra is infinite-dimensional. 
The ground state of the Hamiltonian 
$\hat H_{_1}$  
satisfies the condition 
\begin{equation} 
\hat A\,\vert \Psi_{_0}\rangle = 0 = \hat B_-\,\vert
\Psi_{_0}\rangle\,. 
\label{eqaps0} 
\end{equation} 
The normalized $n$-th excited state of $\hat H_{_1}$ is  
\begin{equation} 
\label{eqpsn} 
\vert \Psi_{_n} \rangle = \frac{1}{\sqrt{R(a_{_1})+ R(a_{_2})+\cdots +
    R(a_n)}} \hat{B}_+ \cdots \frac{1}{\sqrt{R(a_{_1})+  
R(a_{_2})}} \hat{B}_+  \frac{1}{\sqrt{R(a_{_1})}}\hat{B}_+ 
\vert \Psi_{_0} \rangle  
\end{equation} 
with the eigenvalue \ $E_n = \hbar\Omega\,e_n$ where 
\begin{equation} 
e_n =  \sum_{k=1}^n R(a_k)\,. 
\label{eqen} 
\end{equation} 
The action of the $\hat{B}_\pm$ operators on that state given in
Eq. (\ref{eqpsn}) is
\begin{equation} 
\label{eqbpmn} 
\hat{B}_+\vert \Psi_{_n}\rangle = \sqrt{{e}_{n+1}}\;\vert 
\Psi_{_{n+1}}\rangle  
\qquad\qquad {\rm and}\qquad\qquad 
\hat{B}_-\vert \Psi_{_n}\rangle = \sqrt{{e}_{n-1}+R(a_{_0})}\; 
\vert \Psi_{_{n-1}}\rangle\,. 
\end{equation} 
Finally, using Eqs.~(\ref{eqbpmn}) it is easy to show that 
 \begin{equation} 
\label{eqh12n} 
\hat H_{_1}\vert \Psi_{_n}\rangle \equiv \hbar\Omega\left(\hat B_+ 
\hat  B_-\right)\vert \Psi_{_n}\rangle =   
\hbar\Omega\,{e}_{n}\,\vert \Psi_{_n}\rangle  
\quad\quad {\rm and}\quad\quad 
\hat B_-\hat B_+\vert \Psi_{_n}\rangle =  
\left\{{e}_n+R(a_{_0})\right\}\vert \Psi_{_n}\rangle\,. 
\end{equation} 
 
 
\section{Quantum Deformation of Shape-Invariant Systems} 
 
\subsection{$q$-Deformed Creation and Annihilation Operators} 
 
To obtain a quantum deformed version of the shape-invariant systems 
Hamiltonian we introduce the $q$-deformed forms for the creation and  
annihilation operators $\hat B_-^{^{(q)}}$ and $\hat B_+^{^{(q)}}$ in
terms of the usual $\hat B_\pm$ operators as 
\begin{equation} 
\hat B_-^{^{(q)}} \equiv \sqrt{[\hat B_-\hat B_+]_{_q}\over \hat
  B_-\hat  B_+}\, \hat B_- = 
\hat B_-\,\sqrt{[\hat B_+\hat B_-]_{_q}\over \hat B_+\hat B_-} 
\label{eqbqe} 
\end{equation} 
and 
\begin{equation} 
\hat B_+^{^{(q)}} \equiv \left(\hat B_-^{^{(q)}}\right)^\dagger = 
\hat B_+\, \sqrt{[\hat B_-\hat B_+]_{_q}\over \hat B_-\hat B_+} =  
\sqrt{[\hat B_+\hat B_-]_{_q}\over \hat B_+\hat B_-}\,\hat B_+ \,, 
\label{eqbqa} 
\end{equation} 
where we used the definition 
\begin{equation} 
[x]_{_q} \equiv {q^x-q^{-x}\over  q-q^{-1}}\,.
\label{eqnq} 
\end{equation} 
For the limit \ $q\rightarrow 1$ \ it is easy to show that  
\begin{equation} 
\lim_{q\rightarrow 1}\;[x]_{_q}  = x\,, 
\label{eqqlim} 
\end{equation} 
and the postulated relations (\ref{eqbqe}) and (\ref{eqbqa}) tend to
the usual generalized ladder operators. 
For any analytical function $f(x)$ it is easy to show that  
\begin{equation} 
\hat B_\pm f ( \hat B_\mp\hat B_\pm) = 
f ( \hat B_\pm\hat B_\mp) B_\pm. 
\label{eqbqbs1} 
\end{equation} 
In particular we have 
 \begin{equation} 
\hat B_\pm  \left[{q^{\pm\hat B_\mp\hat B_\pm}\over \hat B_\mp\hat
B_\pm}\right]  = \left[{q^{\pm\hat B_\pm\hat B_\mp}\over \hat
B_\pm\hat B_\mp}\right] \hat B_\pm. 
\label{eqbqbs} 
\end{equation} 
 
\subsection{$q$-Deformed Hamiltonian} 
 
With our definition of $q$-deformed ladder operators we can write
down the $q$-deformed form of the Hamiltonian $\hat H_{_1}$ as 
\begin{equation} 
\hat H_{_1}^{^{(q)}} = \hbar\Omega\left(\hat B_+^{^{(q)}}  \hat
B_-^{^{(q)}}\right) =   \hbar\Omega\left(\sqrt{[\hat B_+\hat
B_-]_{_q}\over  \hat B_+\hat B_-}\,\hat B_+  \hat B_-\,\sqrt{[\hat
B_+\hat B_-]_{_q}\over \hat B_+\hat B_-}  \right)=  \hbar\Omega\,[\hat
B_+\hat B_-]_{_q} 
\label{eqh1q} 
\end{equation} 
and also  
\begin{equation} 
\hat B_-^{^{(q)}}\hat B_+^{^{(q)}} =   \sqrt{[\hat B_-\hat
B_+]_{_q}\over \hat B_-\hat B_+}\,\hat B_-  \hat B_+\,\sqrt{[\hat
B_-\hat B_+]_{_q}\over \hat B_-\hat B_+} =  [\hat B_-\hat B_+]_{_q}\,. 
\label{eqh2q} 
\end{equation} 
With these results we conclude that the commutator of the $q$-deformed
$\hat B$'s operators is given by 
\begin{equation} 
[\hat B_-^{^{(q)}},\hat B_+^{^{(q)}}] = [\hat B_-\hat B_+]_{_q} -
[\hat B_+\hat B_-]_{_q}\,. 
\label{eqcbqs} 
\end{equation} 
Another conclusion is that the $q$-deformed version of the Hamiltonian
and its undeformed version commute with each other,   $[\hat
H_{_{1}}^{^{(q)}},\hat H_{_{1}}] = 0$,  therefore them have the common
set of eigenstates give by  Eq.~(\ref{eqpsn}). Then taking into
account Eqs.~(\ref{eqbpmn}), (\ref{eqh12n}), (\ref{eqbqe}), and
(\ref{eqbqa}) we can show that 
\begin{equation} 
\label{eqbqpmn} 
\hat{B}_+^{^{(q)}}\vert \Psi_{_n}\rangle =
\sqrt{[{e}_{n+1}]_{_q}}\;\vert \Psi_{_{n+1}}\rangle   \qquad\quad {\rm
and}\quad\qquad  \hat{B}_-^{^{(q)}}\vert \Psi_{_n}\rangle =
\sqrt{[{e}_{n-1} +  R(a_{_0})]_{_q}}\; \vert \Psi_{_{n-1}}\rangle\,. 
\end{equation} 
To get the eigenvalues of $\hat H_{_{1}}^{^{(q)}}$ we can use these
results and Eqs.~(\ref{eqh1q}) and (\ref{eqh2q}) to obtain 
\begin{equation} 
\label{eqhq12n} 
\hat H_{_1}^{^{(q)}}\vert \Psi_{_n}\rangle =   \hbar\Omega\,\hat
B_+^{^{(q)}}\hat B_-^{^{(q)}}\vert \Psi_{_n}\rangle =
\hbar\Omega\,[{e}_{n}]_{_q}\,\vert \Psi_{_n}\rangle   \quad\quad {\rm
and}\quad\quad  \hat B_-^{^{(q)}}\hat B_+^{^{(q)}}\vert
\Psi_{_n}\rangle =   [{e}_n+R(a_{_0})]_{_q}\,\vert \Psi_{_n}\rangle\,. 
\end{equation} 
 
\subsection{Generalized Quantum Deformed Models for Shape-Invariant 
Potentials} 
 
With the definitions presented in the previous section it is possible
to define new $q$-deformed ladder operators and their $q$-commutations
relations as we illustrate in this section. 
 
\subsubsection{Standard model} 
 
Using Eqs.(\ref{eqh1q}), (\ref{eqh2q}) and the commutation relation
(\ref{eqcb1}) we can evaluate the product 
\begin{eqnarray} 
\hat B_-^{^{(q)}}\hat B_+^{^{(q)}} -   q^{R(a_{_0})}\hat
B_+^{^{(q)}}\hat B_-^{^{(q)}} &=&   [\hat B_-\hat B_+]_{_q} -
q^{R(a_{_0})}[\hat B_+\hat B_-]_{_q}  \nonumber\\  &=& {1\over
q-q^{-1}} \left\{q^{\hat B_-\hat B_+} - q^{-\hat B_-  \hat B_+}-
q^{(\hat B_-\hat B_+-\hat B_+\hat B_-)}\left[q^{\hat B_+\hat B_-} -
q^{-\hat B_+\hat B_-}\right]\right\}\nonumber\\  &=& {q^{-\hat B_+\hat
B_-}\over q-q^{-1}}\left\{q^{(\hat B_-\hat B_+- \hat B_+\hat B_-)}-
q^{(-\hat B_-\hat B_++\hat B_+\hat B_-)}\right\}\nonumber\\ 
\label{eqcom1} 
\end{eqnarray} 
which gives 
\begin{eqnarray} 
\hat B_-^{^{(q)}}\hat B_+^{^{(q)}} -   q^{R(a_{_0})}\hat
B_+^{^{(q)}}\hat B_-^{^{(q)}} &=&  [R(a_{_0})]_{_q}\,q^{-\hat B_+\hat
B_-}\,. 
\label{eqcom2} 
\end{eqnarray} 
In a similar way we can show that 
\begin{eqnarray} 
\hat B_-^{^{(q)}}\hat B_+^{^{(q)}} -   q^{-R(a_{_0})}\hat
B_+^{^{(q)}}\hat B_-^{^{(q)}} &=&   [R(a_{_0})]_{_q}\,q^{\hat B_+\hat
B_-}\,, 
\label{eqcom3} 
\end{eqnarray} 
an expected result considering the invariance under the substitution
of $q\rightarrow q^{-1}$ of the $q$-number definition (\ref{eqnq}).
Eqs.~(\ref{eqcom2}) and (\ref{eqcom3}) represent q-commutators for any
shape-invariant potential. In the particular case of the  harmonic
oscillator potential we have  $R(a_{_0}) = R(a_{_1}) = \dots = {\rm
cte.} = 1$ after a suitable normalization and we get 
\begin{equation} 
\hat B_-^{^{(q)}} \longrightarrow \hat a_{_q}= \sqrt{[\hat N -1]_{_q}
\over \hat N -1}\,\hat a =  \hat a\,\sqrt{[\hat N]_{_q}\over \hat
N}\,,\qquad  \hat B_+^{^{(q)}} \longrightarrow \hat a_{_q}^\dagger =
\hat a^\dagger\,\sqrt{[\hat N -1]_{_q}\over \hat N -1} =  \sqrt{[\hat
N]_{_q}\over \hat N}\,\hat a^\dagger\,. 
\label{eqho} 
\end{equation} 
Therefore, Eqs.~(\ref{eqcom2}) and (\ref{eqcom3}) reduce to the form 
 \begin{equation} 
\hat a_{_q}\hat a_{_q}^\dagger- q^{\pm 1}a_{_q}^\dagger\hat a_{_q}  =
q^{\mp \hat N}\,, 
\label{eqho1} 
\end{equation} 
where \ $\hat N = \hat a^\dagger\hat a \neq \hat a_{_q}^\dagger\hat
a_{_q}$. The operators $\hat a_{_q}$ and $\hat a_{_q}^\dagger$ and its
$q$-deformed commutation relation (\ref{eqho1}) are the basic
assumptions usually postulated in the study of the standard
$q$-deformed harmonic oscillator models
\cite{Bonatsos:1999xj,Macfarlane:dt,Biedenharn:1989jw}. The relation
(\ref{eqho1}) also is termed $q$-deformed  {\it physics} boson
canonical commutation relation \cite{ref25b} and was introduced   in
order to provide a realization of quantum groups
\cite{Macfarlane:dt,Biedenharn:1989jw} which arise   naturally in the
solution of certain lattice models \cite{Batchelor:1989uk}.
 
\subsubsection{Generalized $Q$-Deformed Models} 
 
A second way to construct a $q$-deformed model for a shape-invariant 
potential can be  
obtained if we define the new operators 
 \begin{equation} 
\cases{\hat C^{^{(q)}}_- \equiv {1\over\sqrt{q}}\,\hat B^{^{(q)}}_-\, 
q^{(\hat B_+\hat B_-)/2} =  
{1\over\sqrt{q}}\,q^{(\hat B_-\hat B_+)/2}\,\hat B^{^{(q)}}_-\cr\cr 
\hat C^{^{(q)}}_+ = \hat C^{^{(q)\dagger}}_- =  
{1\over\sqrt{q}}\,q^{(\hat B_+\hat B_-)/2}\,\hat B^{^{(q)}}_+ =  
{1\over\sqrt{q}}\,\hat B^{^{(q)}}_+\,q^{(\hat B_-\hat B_+)/2}\,. 
\cr} 
\label{eqcomm4} 
\end{equation} 
Using the results of Eqs. (\ref{eqh1q}), (\ref{eqh2q}),   the
commutation relation (\ref{eqcb1}) and the commutation between any
function of the remainders $R(a_{_n})$ and the couple of operators
$\hat B_\pm\hat B_\mp$  we can evaluate the products 
\begin{eqnarray} 
\hat C^{^{(q)}}_-\hat C^{^{(q)}}_+ &=&   {1\over\sqrt{q}}\,q^{(\hat
B_-\hat B_+)/2}\,\hat B^{^{(q)}}_-  \hat B^{^{(q)}}_+\,q^{(\hat
B_-\hat B_+)/2} \,  {1\over\sqrt{q}}\;\;\;\;\;\;\;\;\;  \nonumber\\
&=& q^{\hat B_-\hat B_+}\,[\hat B_-\hat B_+]_{_q}/q\nonumber\\  &=&
{q^{2\hat B_-\hat B_+}-1\over q^2-1}\,, 
\label{eqceaqsm} 
\end{eqnarray} 
and  
\begin{eqnarray} 
\hat C^{^{(q)}}_+\hat C^{^{(q)}}_- &=&   {1\over\sqrt{q}}\,q^{(\hat
B_+\hat B_-)/2}\,\hat B^{^{(q)}}_+  \hat B^{^{(q)}}_-\,q^{(\hat
B_+\hat B_-)/2}\,{1\over\sqrt{q}} \;\;\;\;\;\;\;\;\;  \nonumber\\  &=&
q^{\hat B_+\hat B_-}\,[\hat B_+\hat B_-]_{_q}\nonumber\\  &=&
{q^{2\hat B_+\hat B_-}-1\over q^2-1}\,. 
\label{eqcaeqsm} 
\end{eqnarray} 
With the results found in Eqs.~(\ref{eqceaqsm}) and (\ref{eqcaeqsm})
and the commutation relation (\ref{eqcb1}) it is possible to establish
the $q$-deformed commutation relation  
\begin{equation} 
\hat C^{^{(q)}}_-\hat C^{^{(q)}}_+ - q^{2R(a_{_0})}  \hat
C^{^{(q)}}_+\hat C^{^{(q)}}_- = q^{R(a_{_0})}\,[R(a_{_0})]_{_q}/q\,. 
\label{eqcaecomm} 
\end{equation} 
 
The definition of $q$-numbers given by Eq.~(\ref{eqnq}) is not the
only possible one.   There is an alternative definition of called the
$Q$-numbers.  Indeed if we change \ $q^2\longrightarrow Q $ \ and use
$Q$-operators generalization of the $Q$-numbers definition  
\begin{equation} 
[x]_{_Q} = {Q^x-1\over Q-1}\,, 
\label{eqnQ} 
\end{equation} 
it is possible to rewrite the definitions (\ref{eqcomm4}) as  
\begin{equation} 
\hat C_\pm^{^{(q)}} = \hat B^{^{(Q)}}_\pm =  
\sqrt{[(\hat B_\pm\hat B_\mp]_{_Q}\over \hat B_\pm\hat B_\mp}\,\hat 
B_\pm = 
\hat B_\pm\,\sqrt{[\hat B_\mp\hat B_\pm]_{_Q}\over \hat B_\mp\hat 
B_\pm} 
\label{eqbQs} 
\end{equation} 
and show that the $q$-deformed commutation relation (\ref{eqcaecomm})
can be written in its $Q$-deformed version as  
\begin{equation} 
\hat B^{^{(Q)}}_-\hat B^{^{(Q)}}_+ - Q^{R(a_{_0})} 
\hat B^{^{(Q)}}_+\hat B^{^{(Q)}}_- = [R(a_{_0})]_{_Q}\,. 
\label{eqcomBQs} 
\end{equation} 

Again, for a harmonic oscillator potential system we have that \  
$R(a_{_0}) = R(a_{_1}) = \dots = 1$ \ and 
\begin{equation} 
\hat C_-^{^{(q)}} \longrightarrow \hat b_{_Q} =  
\hat a\;\sqrt{[\hat N]_{_Q}\over\hat N} =  
\sqrt{[\hat N-1]_{_Q}\over \hat N-1}\;\hat a\,,\qquad 
\hat C_+^{^{(q)}} \longrightarrow \hat b_{_Q}^\dagger =  
\hat a^\dagger\;\sqrt{[\hat N-1]_{_Q}\over \hat N-1} =  
\sqrt{[\hat N]_{_Q}\over\hat N}\;\hat a^\dagger\,. 
\label{eqhoq1m} 
\end{equation} 
In this case Eq.~(\ref{eqcaecomm}) reduces to the form 
\begin{equation} 
\hat b_{_Q}\hat b_{_Q}^\dagger- Q\,b_{_Q}^\dagger\hat b_{_Q} = 1 
\label{eqho2m} 
\end{equation} 
that together with the $\hat b_{_Q}$'s operators definition correspond
to a   different version of the deformed harmonic oscillator model,
first introduced by Arik and Coon \cite{Arik:1973vg} and later
considered also by  Kuryshkin \cite{ref27}. Thus, we can
consider the operators $\hat B_\pm^{^{(Q)}}$ and Eq.~(\ref{eqcomBQs})
as the generalized version for all shape-invariant systems of the
$Q$-deformed basic relations  first postulated by Arik and Coon for
the $Q$-deformed harmonic oscillator model. 
 
\subsubsection{Another generalized $q$-deformed model} 
 
Another $q$-deformed model can be obtained if we define the
new operators  
\begin{equation} 
\cases{\hat D^{^{(q)}}_- = q^{-R(a_{_0})/2}\,\hat B^{^{(q)}}_-\,q^{ 
\hat B_+\hat B_-/2} =  
q^{-R(a_{_0})/2}\,q^{\hat B_-\hat B_+/2}\,\hat B^{^{(q)}}_-\cr\cr 
\hat D^{^{(q)}}_+ = \hat D^{^{(q)\dagger}}_- =  
q^{\hat B_+\hat B_-/2}\,\hat B^{^{(q)}}_+\,q^{-R(a_{_0})/2} =  
\hat B^{^{(q)}}_+\,q^{\hat B_-\hat B_+/2}\,q^{-R(a_{_0})/2}\,.  
\cr} 
\label{eqcom4} 
\end{equation} 
Using the results of Eqs.(\ref{eqh1q}), (\ref{eqh2q}), the commutation
relation (\ref{eqcb1}) and the commutation between any function of the
remainders $R(a_{_n})$ and the couple of operators $\hat B_\pm\hat
B_\mp$ we can write down the product
\begin{eqnarray} 
\hat D^{^{(q)}}_-\hat D^{^{(q)}}_+ &=&  
q^{-R(a_{_0})/2}\,q^{\hat B_-\hat B_+/2}\,\hat B^{^{(q)}}_- 
\hat B^{^{(q)}}_+\,q^{\hat B_-\hat B_+/2}\,q^{-R(a_{_0})/2}
\;\;\;\;\;\;\;\;\; 
\nonumber\\ 
&=& q^{-R(a_{_0})}\,q^{\hat B_-\hat B_+}\,[\hat B_-\hat B_+]_{_q}
\nonumber\\ 
&=& {q^{(\hat B_-\hat B_++\hat B_+\hat B_-)}- 
q^{-R(a_{_0})}\over q-q^{-1}}\,, 
\label{eqceaqs} 
\end{eqnarray} 
and, with the help of Eqs.~(\ref{eqrbpm}), the product 
\begin{eqnarray} 
\hat D^{^{(q)}}_+\hat D^{^{(q)}}_- &=&  
q^{\hat B_+\hat B_-/2}\,\hat B^{^{(q)}}_+q^{-R(a_{_0})} 
\hat B^{^{(q)}}_-\,q^{\hat B_+\hat B_-/2}\nonumber\\ 
&=& q^{-R(a_{_1})}\,q^{\hat B_+\hat B_-}\,[\hat B_+\hat B_-]_{_q}
\nonumber\\ 
&=& q^{-[R(a_{_0})+R(a_{_1})]}\left({q^{(\hat B_+\hat B_- +
\hat B_-\hat B_+)}-q^{R(a_{_0})} 
\over q-q^{-1}}\right)\,. 
\label{eqcaeqs} 
\end{eqnarray} 
Using the results shown in Eqs.~(\ref{eqceaqs}) and (\ref{eqcaeqs}) it
is possible to establish the following $q$-deformed commutation
relation 
\begin{equation} 
\hat D^{^{(q)}}_-\hat D^{^{(q)}}_+ - q^{[R(a_{_0})+R(a_{_1})]} 
\hat D^{^{(q)}}_+\hat D^{^{(q)}}_- = [R(a_{_0})]_{_q}\,. 
\label{eqcaecom} 
\end{equation} 
In the limiting case of a harmonic oscillator potential system, when  
$R(a_{_0}) = R(a_{_1}) = \dots = 1$ one has   
\begin{equation} 
\hat D_-^{^{(q)}} \longrightarrow \hat b_{_q} = q^{-1/2}\hat 
a_{_q}q^{\hat N/2} =  
q^{\hat N/2}\hat a_{_q}\,,\qquad 
\hat D_+^{^{(q)}} \longrightarrow \hat b_{_q}^\dagger =  
q^{-1/2}q^{\hat N/2}\hat a_{_q}^\dagger = \hat a_{_q}^\dagger q^{\hat N/2}\,, 
\label{eqhoq1} 
\end{equation} 
which gives us for Eq.~(\ref{eqcaecom}) the following form: 
\begin{equation} 
\hat b_{_q}\hat b_{_q}^\dagger- q^2b_{_q}^\dagger\hat b_{_q} = 1\,. 
\label{eqho2} 
\end{equation} 
Hence there are two possible and different shape invariant
generalizations for Arik and Coon quantum deformed model. The first
one by using the operators $\hat B_\pm^{^{(Q)}}$    and
Eq.~(\ref{eqcomBQs}) and the second one with the operators $\hat
D_\pm^{^{(q)}}$ and   Eq.~(\ref{eqcaecom}). These two generalizations
for shape invariant systems are equivalent   when we applied to the
harmonic oscillator potential system, giving the standard Arik and
Coon model.  
 
An important aspect to observe at this point is that all quantum
deformed models generalized   from the primary shape-invariant
potentials and presented in these last sections do not preserve   the
shape invariance after the quantum deformation. In other words, the
quantum deformation breaks   the shape invariance of the final
$q$-deformed system. Obviously this fact is a result of the basic
assumptions used to build the quantum deformed models.  
 
 
\section{Shape-Invariant Quantum Deformed Systems}
 
The purpose of this section is to build an alternative generalized
quantum   deformed model which, unlike the previous ones, after the
quantum deformation, preserves the shape invariance
condition shown by the primary system.

\subsection{New $q$-Deformed  Ladder Operators} 
 
To find a $q$-deformed system formulation which preserves the
shape-invariant condition we introduce new operators defined by 
\begin{equation} 
\cases{\hat {S}_-^{^{(q)}} = {\cal F} 
\,\hat B_-^{^{(q)}}\,q^{\hat B_+\hat B_-/2} =  
{\cal F}\,q^{\hat B_-\hat B_+/2}\,\hat B_-^{^{(q)}}\cr\cr 
\hat {S}_+^{^{(q)}} = \hat {S}_-^{^{(q)\dagger}} = 
q^{\hat B_+\hat B_-/2}\,\hat B_+^{^{(q)}} 
\,{\cal F} = \hat B_+^{^{(q)}}\,q^{\hat B_-\hat B_+/2}\,{\cal F}\cr} 
\label{eqdop} 
\end{equation} 
where the $\hat B_\pm^{^{(q)}}$ operators were introduced by
Eqs.(\ref{eqbqa}), (\ref{eqbqe}) and ${\cal F}$ is a compact notation
for a real functional of the potential parameters $a_0, a_1, a_2,
\dots$. Note that for the hermitian conjugation condition written
above to be satisfied $q$ must be assumed as a real parameter.  We
specify the conditions on ${\cal F}$ below.  Considering that 
$\left[{\cal F}, \hat B_\pm\hat B_\mp\right] = 0$ and using the
definitions in Eq. (\ref{eqdop}), the commutation relations
(\ref{eqcb1}) and   Eqs.~(\ref{eqh1q}), (\ref{eqh2q}) and
(\ref{eqrbpm}) it is possible to evaluate the   products 
\begin{eqnarray} 
\hat {S}_-^{^{(q)}}\hat {S}_+^{^{(q)}} &=& 
{\cal F}\,q^{\hat B_- \hat B_+/2}\,\hat B_-^{^{(q)}} 
\hat B_+^{^{(q)}}\,q^{\hat B_-\hat B_+/2}\,{\cal F} \nonumber\\ 
&=& {\cal F}^2 \,q^{\hat B_-\hat B_+}\,[\hat B_-\hat B_+]_{_q} 
\label{eqdeda} 
\end{eqnarray} 
and 
\begin{eqnarray} 
\hat {S}_+^{^{(q)}}\hat {S}_-^{^{(q)}} &=& 
q^{\hat B_+\hat B_-/2}\,\hat B_+^{^{(q)}}\,{\cal F}^2
\,\hat B_-^{^{(q)}}\,q^{\hat B_+\hat B_-/2}\;\;\;\;\;\;\;\;\;
\;\;\;\nonumber\\ 
&=& \hat T(a_1)\,{\cal F}^2 \hat T^\dagger(a_1) \,q^{\hat B_ +  \hat
  B_-} \,[\hat B_+\hat B_-]_{_q}\,. 
\label{eqdade} 
\end{eqnarray} 
Now, with the results of Eqs. (\ref{eqdeda}), (\ref{eqdade}) and the
commutation relation (\ref{eqcb1}) we can write down the commutator of
the $\hat {S}$-operators  
\begin{equation} 
\left[\hat {S}_-^{^{(q)}},\hat {S}_+^{^{(q)}}\right] =  
{\cal F}^2 \,q^{R(a_{_0})}\,q^{\hat B_+\hat B_-}\, [\hat B_-\hat
B_+]_{_q} - 
\hat T(a_1)\, {\cal F}^2 \hat T^\dagger(a_1)  \,q^{-R(a_{_0})} \,
q^{\hat B_-\hat B_+}\, [\hat B_+\hat B_-]_{_q}\,. 
\label{eqdedac} 
\end{equation} 
At this point, we assume that the functional operator ${\cal F}$,
until now considered arbitrary, satisfies the constraint 
\begin{equation} 
\hat T(a_1)\, {\cal F}^2 T^\dagger(a_1)  = q^{2 R(a_{_0})}\,{\cal
  F}^2 \,. 
\label{eqfvin} 
\end{equation} 
Thus, taking into account this condition and the operator relation  
\begin{equation} 
q^{\hat B_\pm\hat B_\mp}\left[\hat B_\mp\hat B_\pm\right]_{_q} =  
{q^{(\hat B_+\hat B_-+\hat B_-\hat B_+)} - q^{\mp R(a_{_0})} 
\over q - q^{-1}}\,, 
\label{eqqbs} 
\end{equation}  
it follows that the commutator can be written as  
\begin{equation} 
\left[\hat {S}_-^{^{(q)}},\hat {S}_+^{^{(q)}}\right] = {\cal
  G}_0 \,,  
\quad\qquad {\rm with}\quad\qquad {\cal G}_0 \equiv 
{\cal F}^2 \,q^{R(a_{_0})}\,\left[R(a_{_0})\right]_{_q}\,. 
\label{eqdeda2} 
\end{equation} 
Comparing Eqs.~(\ref{eqcb1}) and (\ref{eqdeda2}) we conclude that the 
later can be associated with a shape-invariance condition as the 
former and that $\hat {S}_-^{^{(q)}}$ and $\hat {S}_+^{^{(q)}}$ 
are the appropriate creation and annihilation operators for the spectra 
of the $q$-deformed shape-invariant systems whose Hamiltonian and its 
eigenstates and eigenvalues will be determined at next section.

\subsection{Hamiltonian, Eigenstates and Eigenvalues} 
 
Using the new ladder operators introduced in the previous  
section we can define a new Hamiltonian as 
\begin{equation} 
\hat {\cal H}^{^{(q)}} = \hbar\Omega\,\hat {S}_+^{^{(q)}} 
\hat {S}_-^{^{(q)}}\,. 
\label{eqhdd} 
\end{equation} 
With this definition, the relations (\ref{eqrbpm}) and
Eq.~(\ref{eqdeda2}) we can write down the additional commutation
relations  
\begin{equation} 
[{\hat {\cal H}}^{^{(q)}},({\hat {S}_+^{^{(q)}}})^n] =  
\hbar\Omega\,\left\{{\cal G}_1 + {\cal G}_2 + \cdots + {\cal G}_n
\right\} \,({\hat {S}_+^{^{(q)}}})^n\;\;\;\;\;  
\label{eqhdac} 
\end{equation} 
and 
\begin{equation} 
[{\hat {\cal H}}^{^{(q)}},({\hat {S}_-^{^{(q)}}})^n] =  
-\hbar\Omega\,({\hat {S}_-^{^{(q)}}})^n\, 
\left\{{\cal G}_1 + {\cal G}_2 + \cdots  +  
{\cal G}_n \right\} 
\label{eqhdec} 
\end{equation} 
where we defined 
\begin{equation} 
{\cal G}_n  =  T(a_1)\, {\cal G}_{n-1} T^\dagger(a_1) ,
\label{eqfggen} 
\end{equation} 
with ${\cal G}_0$ given by Eq. (\ref{eqdeda2}). 
Using Eqs.~(\ref{eqaps0}), (\ref{eqbqe}) and (\ref{eqdop})  
we can also show that \ $\hat {S}_-^{^{(q)}}\,\vert \Psi_{_0}\rangle =
0\,$. From this result and the commutator (\ref{eqhdac}) it follows
that  
\begin{equation} 
{\hat {\cal H}}^{^{(q)}}\{({\hat {S}_+^{^{(q)}}})^n\, \vert
\Psi_{_0}\rangle\} =  \hbar\Omega \, \left\{{\cal G}_1 +
  {\cal G}_2 + \cdots +  {\cal G}_n \right\} \, \{({\hat
  {S}_+^{^{(q)}}})^n\,\vert \Psi_{_0}\rangle\}\,,  
\label{eqhdac1} 
\end{equation} 
i.e., \ $({\hat {S}_+^{^{(q)}}})^n\,\vert \Psi_{_0}\rangle$ is an
eigenstate of the Hamiltonian ${\hat {\cal H}}^{^{(q)}}$ with the
eigenvalue   
\begin{equation} 
{\cal E}_n = \hbar\Omega\,\sum_{k=1}^n\, {\cal G}_k \,. 
\label{eqeqhn} 
\end{equation} 
Indeed this conclusion could be obtained with another approach. With
the operator definition in Eq. (\ref{eqdop}) and the functional
condition in Eq. (\ref{eqfvin}), the Hamiltonian ${\hat {\cal
    H}}^{^{(q)}}$ can be written in terms of the primary
shape-invariant operators $\hat B_\pm$ as   
\begin{equation} 
\hat {\cal H}^{^{(q)}} = \hbar\Omega\,q^{2R(a_{_0})} \, {\cal F}^2 \, 
q^{\hat B_+\hat B_-}\,[\hat B_+\hat B_-]_{_q}\,. 
\label{eqhddb} 
\end{equation} 
Obviously we have that \ $[\hat {\cal H}^{^{(q)}},\hat H_{_{1}}] = 0$,
hence these two Hamiltonians have a common set of eigenstates give by
Eq.~(\ref{eqpsn}), 
in other words \ $\vert \Psi_{_n}\rangle \propto  
({\hat {S}_+^{^{(q)}}})^n\,\vert \Psi_{_0}\rangle$.  
Now, to get the eigenvalues of $\hat {\cal H}^{^{(q)}}$ it is enough
to use this expression, Eqs.~(\ref{eqh12n}) and (\ref{eqhq12n}) to
obtain   
\begin{equation} 
\hat {\cal H}^{^{(q)}}\vert \Psi_{_n}\rangle =  
\hbar\Omega\,q^{2R(a_{_0})}\,{\cal F}^2\, 
q^{{e}_n}\,[{e}_{n}]_{_q}\,\vert \Psi_{_n}\rangle 
\label{eqhqen} 
\end{equation} 
where ${e}_n$ is given by Eq. (\ref{eqen}). Indeed, by using the
generalization of the condition in Eq. (\ref{eqfvin})  
\begin{equation} 
\{\hat T(a_{_1})\}^k\, {\cal F}\,  
\{\hat T^\dagger(a_{_1})\}^k = \prod_{j=0}^{k-1}\,q^{R(a_{_j})}\,{\cal
  F} \label{eqving} 
\end{equation} 
it is straightforward to show that 
\begin{equation} 
{\cal E}_n = \hbar\Omega\,\sum_{k=1}^n\, {\cal G}_k =  
\hbar\Omega\,\sum_{k=1}^n\, \{\hat T(a_{_1})\}^k\, {\cal F}\,  
\{\hat T^\dagger(a_{_1})\}^k  
\,q^{R(a_{_k})}\,\left[R(a_{_k})\right]_{_q} =  
\hbar\Omega\,q^{2R(a_{_0})}\,{\cal F}^2 \,q^{{e}_n}\, 
[{e}_{n}]_{_q}\,.
\label{eqeqhn3} 
\end{equation} 
On the other hand, with Eqs.~(\ref{eqrbpm}), (\ref{eqh12n}),
(\ref{eqbqpmn}), (\ref{eqdop}) and the condition (\ref{eqfvin}) it is
possible to show that  
\begin{equation} 
\label{eqdqpa} 
\hat{S}_+^{^{(q)}}\vert \Psi_{_n}\rangle = q^{R(a_{_0})}\,{\cal F} 
\,q^{{e}_{n+1}/2}\,\sqrt{[{e}_{n+1}]_{_q}}\;\vert \Psi_{_{n+1}}\rangle 
\;\;\;\;\;\;\;\;\;\;\;\;\;\;\;\;\; 
\end{equation} 
and 
\begin{equation} 
\label{eqdqpe} 
\hat{S}_-^{^{(q)}}\vert \Psi_{_n}\rangle = q^{R(a_{_0})/2}\,{\cal F} 
\,q^{{e}_{n-1}/2}\,\sqrt{[{e}_{n-1}+R(a_{_0})]_{_q}}\; 
\vert \Psi_{_{n-1}}\rangle\,, 
\end{equation} 
showing clearly that $\hat{S}_+^{^{(q)}}$ and $\hat{S}_-^{^{(q)}}$  
represent appropriate creation and annihilation operators. 
 
In addition to the commutation relations in Eqs. (\ref{eqhdac})
and (\ref{eqhdec}) we can establish the commutation relations 
\begin{equation} 
\left[\hat {S}_+^{^{(q)}},{\cal G}_j \right] =  
\{{\cal G}_{j+1} - {\cal G}_j \}\,\hat {S}_+^{^{(q)}}\,, 
\label{eqdr0com} 
\end{equation} 
\begin{equation} 
\left[\hat {S}_+^{^{(q)}},\{{\cal G}_{j+1} - {\cal G}_j \}\right] =  
\{{\cal G}_{j+2} - {\cal G}_j \} \, (\hat {S}_+^{^{(q)}})^2\,, 
\label{eqdr1com} 
\end{equation} 
and so on. In general, there is an infinite number of these
commutation relations that with their complex conjugates together with
Eq.~(\ref{eqdeda2}) form an infinite-dimensional Lie algebra, realized
here in an unitary representation.  
 
\subsection{An Ansatz for the Functional ${\cal F}$} 
 
Until this point we did not specify the functional ${\cal
F}$. In this section we will give an ansatz for ${\cal
F}$. We will begin by noting that the remainder $R(a_j)$ is given by
the expression
\begin{equation}
  \label{eq:aa1}
  R( a_j ) = C \left[ f( a_j)  -  f(a_{j+1}) \right]
\end{equation}
for a number of shape-invariant potentials. In Eq. (\ref{eq:aa1}) both
the constant $C$ and the function $f( a_j)$ is determined by the
particular potential in consideration.

For the Morse potential, $V(x) = V_0 ( e^{-2\lambda x} - 2 b
  e^{-\lambda 
  x})$, the superpotential is \cite{Balantekin:1997mg} 
\begin{equation}
  \label{eq:aa2}
  W(x; a_n) =  \sqrt{V_0} ( a_n - e^{- \lambda x}).
\end{equation}
$R(a_j)$ is given by 
\begin{equation}
  \label{eq:aa3}
  R( a_j ) =  \left[ a_j^2  -  a_{j+1}^2 \right]
\end{equation}
with 
\begin{equation}
  \label{eq:aa4}
  a_j = b -  \frac{\lambda\hbar}{\sqrt{2mV_0}} (j -\frac{1}{2}),
\end{equation}
where we identified $\hbar \Omega \equiv V_0$. Hence we get $f( a_j) = a_j^2$.

For the Scarf potential, $V(x) =-V_0/ \cosh^2 \lambda x$, the
superpotential is \cite{Balantekin:1997mg} 
\begin{equation}
  \label{eq:aa5}
   W(x; a_n) =  \sqrt{V_0}  a_n \tanh \lambda x . 
\end{equation}
$R(a_j)$ is given by 
\begin{equation}
  \label{eq:aa6}
  R(a_j) = \left[ a_j^2  -  a_{j+1}^2
  \right] 
\end{equation}
with 
\begin{equation}
  \label{eq:aa7}
  a_j = \frac{1}{2} \left( \sqrt{\frac{8mV_0}{\hbar^2\lambda^2}+1}
    -2j+1 \right) 
\end{equation}
if we identify $\Omega = \lambda \sqrt{V_0/(2m)}$. 
Hence one again obtains $f( a_j) = a_j^2$. 

For the Coulomb potential, $V_L (r) = - \frac{Ze^2}{r} +
\frac{\hbar^2 \lambda^2}{2m} L(L+1)$, we have $a_L = L$. The
superpotential is \cite{DrigoFilho:2001ae} 
\begin{equation}
    \label{eq:aa8}
W(r; L) = \sqrt{\frac{m (Ze^2)^2}{2 \hbar^2}} \left( 
\frac{1}{(L+1)} - \frac{\hbar^2}{mZe^2} \frac{L+1}{r} \right) 
\end{equation}
with 
\begin{equation}
  \label{eq:aa9}
R( L ) = \frac{Z^2 e^4}{4} \left[ \frac{1}{(L+1)^2}
  - \frac{1}{(L+2)^2} \right] ,
\end{equation}
if we identify $\hbar \Omega = m (Ze^2)^2/(2 \hbar^2)$. 
In this case $f(L) =   \frac{1}{(L+1)^2}$ or alternatively $f( a_j) =
\frac{1}{(a_j+1)^2}$. 

We can give an explicit ansatz for ${\cal F}$ applicable to those cases
where Eq. (\ref{eq:aa1}) is satisfied. Noting the identity
\begin{equation}
  \label{eq:aa10}
 \hat T(a_{_1})\, q^{-2 C f(a_0)} \, \hat T^\dagger(a_{_1}) =  q^{-2 C
 f(a_1)} = q^{2R(a_0)}  q^{-2 C f(a_0)} ,
\end{equation}
we conclude that in those cases the functional ${\cal F}$ depends only
on $a_0$: 
\begin{equation}
  \label{eq:aa11}
  {\cal F} =  q^{- C f(a_0)} . 
\end{equation}
For the shape-invariant potentials that satisfy Eq. (\ref{eq:aa1}) we
can then define
\begin{equation}
  \label{eq:aa12}
  {\cal F}_0 \equiv  q^{- C f(a_0)} ,
\end{equation}
and 
\begin{equation} 
{\cal F}_n  =  T(a_1)\, {\cal F}_{n-1} T^\dagger(a_1) ,
\label{none} 
\end{equation} 
At the moment a more general expression for ${\cal F}$  is not readily
available. 

 
\section{Conclusions} 
 
In this article we introduced a quantum deformed theory applicable to
all shape-invariant systems. To achieve this we introduced the
appropriate $q$-deformed ladder operators. We also constructed an
alternative $q$-deformed model that preserves the shape-invariance
property presented by primary system.  Our results are applicable to
those shape-invariant potentials where the potential parameters are
related by a translation.

Shape-invariance represents exact-solvability of a system.  
We previously given a method to obtain a new exactly-solvable system
starting with a known shape-invariant system by coupling it to a
two-level system. The resulting models generalize the Jaynes-Cummings
model by substituting a shape-invariant system instead of the harmonic
oscillator \cite{Aleixo:2000ub,Aleixo:2001jm}. 
Results presented in this paper represent generalization in a
different direction, namely deformation of not the harmonic
oscillator, but of certain shape-invariant
systems.

\section*{ACKNOWLEDGMENTS} 
 
This  work   was supported in  part  by   the  U.S.  National  Science
Foundation Grants No.\ INT-0070889  and PHY-0244384  at the
University  of  Wisconsin, and  in  part by  the  University of
Wisconsin Research  Committee   with  funds  granted by    the
Wisconsin Alumni  Research  Foundation.  M.A.C.R.\ acknowledges  the
support of Funda\c  c\~ao de  Amparo \`a Pesquisa  do Estado de  S\~ao
Paulo - FAPESP (Contract No.\  98/13722-2).    A.N.F.A. and M.A.C.R.
acknowledge  the  support  of  Conselho     Nacional      de
Pesquisa        (CNPq)       (Contract  No. 910040/99-0) and thank to
the Nuclear Theory Group at University   of Wisconsin for their very
kind hospitality. 
  

\end{document}